\documentclass[seceq]{ptptex}
\newcommand{\D}{\displaystyle}

\usepackage{epsfig}
\usepackage{graphicx}

\title{%        %You can use \\ for explicit line-break
Proton-rich Nuclei at and beyond the Proton Drip Line in the
Relativistic Mean Field Theory}
\author{%       %Use \sc for the family name
L. S. {\sc Geng},$^{1,3,}$\footnote{E-mail:
lsgeng0@rcnp.osaka-u.ac.jp} H. {\sc
Toki}$^{1,2,}$\footnote{E-mail: toki@rcnp.osaka-u.ac.jp} and J.
{\sc Meng}$^{3,}$ \footnote{E-mail: mengj@pku.edu.cn} }
\inst{%         %Affiliation, neglected when [addenda] or [errata]
$^1$Research Center for Nuclear Physics (RCNP), Osaka University,\\
Ibaraki 567-0047, Japan
\\
$^2$The Institute of Physical and Chemical Research (RIKEN),\\
Wako 351-0198, Japan\\
$^3$School of Physics, Peking University, Beijing 100871, P. R.
China }
\abst{%       %this abstract is neglected when [addenda] or [errata]
Ground state properties of proton-rich odd-$Z$ nuclei in the
region $55\le Z \le 73$ are studied in the relativistic mean field
(RMF) theory. The RMF equations are solved by using the expansion
method in the Harmonic-Oscillator basis. In the particle-particle
channel, we use the state-dependent BCS method with a zero-range
$\delta$-force, which has been proved to be effective even for
neutron-rich nuclei. All the ground state properties, including
the one-proton separation energies, the ground state deformations,
the last occupied proton orbits and the locations of proton drip
line, are calculated. Good agreement with both the available
experimental data and the predictions of the RHB method are
obtained.}

\begin{document}
\maketitle
\section {Introduction}
The structure and decay modes of nuclei at and beyond the proton
drip line represent one of the most active areas in both
experimental and theoretical studies of exotic nuclei with extreme
isospin ratios. Recent reviews can be found in Refs.
\cite{woods.97,aberg.97} and references therein. For each element,
the number of neutrons that can be supported by the nucleus is
limited if the nucleus is to remain stable with respect to
nucleonic emission. These limits define the proton and neutron
drip lines. The main difference between the proton and the neutron
is the existence of the Coulomb potential for the former.
Consequently, the proton drip line lies much closer to the
stability line than the neutron drip line. For example, the
lightest particle-stable neon isotope is $^{17}_{10}$Ne$_{7}$
\cite{woods.97} , while as heavy an isotope as
$^{34}_{10}$Ne$_{24}$ \cite{notani.02} is known to exist. The
existence of the Coulomb barrier also traps the wave function of
the proton in the nuclear region, which explains why we have not
yet clearly observed halo phenomena in proton-rich nuclei.

The proton drip line has been almost fully mapped up to $Z=21$
\cite{woods.97} experimentally. No proton-unbound nuclei have been
directly observed in this region, which is not surprising, since
the Coulomb barrier is relatively low.  It has been argued that
ground state proton decay can only be detected directly for nuclei
with $Z>50$, where the relatively high potential energy barrier
causes nuclei to survive long enough to be detected experimentally
\cite{woods.97}. Detailed studies of ground state proton
radioactivity have been reported for odd-$Z$ nuclei mainly in the
two mass regions, $51\le Z\le 55$ and $69\le Z \le 83$. The
experimental features observed for most of these nuclei have been
explained by assuming them to have spherical shapes
\cite{aberg.97}. The half-lives for proton decay were then
evaluated using the semi-classical WKB method or standard reaction
theory within the distorted wave approximation.  However, to
obtain proton decay rates in the transitional region beyond the
$Z=50$ shell closure ($^{109}$I and $^{112,113}$Cs) and in the
region of light rare-earth nuclei ($^{131}$Eu and $^{140,141}$Ho)
requires calculations assuming significant prolate deformations.
Recently, an exact formalism was developed in Refs. \cite{magl.98,
magl.99}, where the half-lives are evaluated with the assumption
that the emitted proton exists in a deformed single particle
Nilsson level. This formalism was applied \cite{magl.98,
magl.99,ferr.00, magl.00} to analyze all data presently available
for odd proton emitters with an even number of neutrons and the
known odd-odd proton emitters $^{112}$Cs, $^{140}$Ho and
$^{150}$Lu \cite{ferr.01}. This formalism, however, does not
predict proton separation energies, i.e. the model does not
predict which nuclei are likely to be proton emitters.

In the past decade, the relativistic mean field (RMF) theory  has
been successfully applied to the study of nuclear properties
throughout the Periodic Table. General reviews of the RMF theory
and its applications in nuclear physics can be found in Refs.
\cite{serot.86,reinhard.89,ring.96,hirata.91}. To study exotic
nuclei whose Fermi surfaces are close to the threshold, several
issues, including the continuum and the pairing correlation, must
be treated carefully \cite{meng.96, meng.98, meng.99, meng.02,
yadav.02, sand.03,sand.00, geng.03}. Luckily, in the case of
proton-rich nuclei, because the nuclear wave function is well
confined in the nuclear region due to the existence of the Coulomb
potential, one only needs to take care of the pairing correlation.
The use of a zero-range $\delta$-force in the particle-particle
channel, $V=-V_0\delta({\vec r_1}-{\vec r_2})$, has proved to be
an efficient and economical way to accomplish this \cite{meng.96,
meng.98, meng.99, meng.02, yadav.02, sand.03,sand.00, geng.03}.
(For detailed discussions of more general cases, including
neutron-rich nuclei, we refer the reader to Refs. $^{13)-20)}$ and
references therein.) In our present work, the deformed RMF+BCS
method with a zero-range $\delta$-force in the pairing channel
\cite{geng.03,geng.031,geng.032} is applied to the analysis of
deformed odd-$Z$ proton-rich nuclei with $55\le Z\le 73$. In the
mean-field part, two of the most successful parameter sets, TMA
\cite{sugahara.94} and NL3, \cite{ring.97} are used.

The RMF theory was first applied to study deformed proton emitters
by Lalazissis et al. In Refs. \cite{lalaprc.99, lalanpa.99,
vret.99, lalanpa.01}, the proton drip-line nuclei with $31\le Z\le
49$ and $51\le Z \le 73$ are studied within the framework of the
Relativistic Hartree-Bogoliubov (RHB) model. In the mean-field
part, the NL3 \cite{ring.97} parameter set is used. In the
particle-particle channel, the pairing part of the Gogny force
with the D1S set \cite{gogny.84} has been adopted. The calculated
one-proton separation energy and other relevant properties are in
good agreement with the available experimental data. Such a
success shows that the RMF theory could be an appropriate method
even for the description of nuclei with extreme isospin ratios.
This observation also motivates us to study deformed proton
emitters within a different framework, that of the deformed
RMF+BCS method, due to the following considerations. First, we
would like to see whether such a success is dependent on the
choice of a particular combination of parameter sets, i.e.
NL3+D1S. Second, we want to see whether the argument that a
zero-range $\delta$-force is an efficient and economical
interaction to treat the pairing correlation holds in such extreme
cases, i.e. at and/or near the proton drip line. Finally, many new
theoretical and experimental works, including those reported in
Refs. \cite {ferr.00,ferr.01,mahm.01,sora.01}, to name just a few,
have been carried out since the last series of works
\cite{lalaprc.99, lalanpa.99, vret.99, lalanpa.01} in the
relativistic mean field theory. Because in general, the BCS method
is thought of as an approximation to the full Bogoliubov
transformation, we would like to compare our results with the
existing RHB results.

The article is organized as follows. In \S2, we give a brief
introduction to the deformed RMF+BCS method. In \S3, we present
our results for odd-$Z$ proton-rich nuclei with $55\le Z \le 73$
and compare our results with the available experimental data and
those of the RHB method. \S4 is devoted to the summary of this
paper.

\section{The deformed RMF+BCS method}

Our RMF calculations have been carried out using the model
Lagrangian density with nonlinear terms both for the $\sigma$ and
$\omega$ mesons, as described in detail in Ref. \cite{geng.03}.
The Lagrangian density is given by

\begin{eqnarray}
{\cal L} &=& \bar \psi (i\gamma^\mu\partial_\mu -M) \psi +
\,\frac{\D 1}{\D 2}\partial_\mu\sigma\partial^\mu\sigma-\frac{\D
1}{\D 2}m_{\sigma}^{2} \sigma^{2}- \frac{\D 1}{ \D
3}g_{2}\sigma^{3}-\frac{\D 1}{\D
4}g_{3}\sigma^{4}-g_{\sigma}\bar\psi
\sigma \psi\nonumber\\
&&-\frac{\D 1}{\D 4}\Omega_{\mu\nu}\Omega^{\mu\nu}+\frac{\D 1}{\D
2}m_\omega^2\omega_\mu\omega^\mu +\frac{\D 1}{\D
4}g_4(\omega_\mu\omega^\mu)^2-g_{\omega}\bar\psi
\gamma^\mu \psi\omega_\mu\nonumber\\
 && -\frac{\D 1}{\D 4}{R^a}_{\mu\nu}{R^a}^{\mu\nu} +
 \frac{\D 1}{\D 2}m_{\rho}^{2}
 \rho^a_{\mu}\rho^{a\mu}
%{\bf {\rho}}_{\mu}{\bf {\rho}}^{\mu}
     -g_{\rho}\bar\psi\gamma_\mu\tau^a \psi\rho^{\mu a}\nonumber \\
      && -\frac{\D 1}{\D 4}F_{\mu\nu}F^{\mu\nu} -e \bar\psi
      \gamma_\mu\frac{\D 1-\tau_3}{\D 2}A^\mu
      \psi,
\end{eqnarray}

where the field tensors of the vector mesons and of the
electromagnetic field take the forms
\begin{equation}
\left\{
\begin {array}{rl}
\Omega_{\mu\nu} =&
\partial_{\mu}\omega_{\nu}-\partial_{\nu}\omega_{\mu},\\
 R^a_{\mu\nu} =& \partial_{\mu}
                  \rho^a_{\nu}
                  -\partial_{\nu}
                  \rho^a_{\mu}-2g_\rho\epsilon^{abc}\rho^b_\mu\rho^c_\nu,\\
 F_{\mu\nu} =& \partial_{\mu}A_{\nu}-\partial_{\nu}
A_{\mu},
\end{array}\right.
\end{equation}
and other symbols have their usual meanings. Based on the
single-particle spectra calculated with the RMF method, we perform
a state-dependent BCS calculation \cite{lane.64,ring.80}. The gap
equation has a standard form for all the single particle states,
\begin{equation}\label{eq:bcs}
\Delta_k=-\frac{\D 1}{\D 2}\sum_{k'>0}\frac{\D \bar{V}_{kk'}
\Delta_{k'}}{\D\sqrt{(\varepsilon_{k'}-\lambda)^2+\Delta_{k'}^2}},
\end{equation}
where $\varepsilon_{k'}$ is the single particle energy and
$\lambda$ is the Fermi energy, whereas the particle number
condition is given by $2\sum\limits_{k>0} v_k^2=N$. In the present
work, for the pairing interaction we use a zero-range
$\delta$-force,
\begin{equation}
V=-V_0\delta({\vec r_1}-{\vec r_2}),
\end{equation}
with the same strength $V_0$ for both protons and neutrons. The
pairing matrix element for the zero-range $\delta$-force is given
by
\begin{equation}\label{eq:me}
\begin{array} {lll}\bar{V}_{ij}&=&\langle i\bar{i}|V|j\bar{j}\rangle-\langle i\bar{i}|V|\bar{j}j\rangle=-V_0\int d^3r\,
\left[\psi^\dagger_i\psi^\dagger_{\bar{i}}\psi_j\psi_{\bar{j}}-\psi^\dagger_i\psi^\dagger_{\bar{i}}\psi_{\bar{j}}\psi_j\right],\\
\end{array}
\end{equation}
with the nucleon wave function in the form
\begin{equation}\label{eq:spinor}
\psi_i({\vec r},t)=\left(\begin{array}{c}f_i({\vec r})\\ig_i({\vec
r})\end{array}\right) =\frac{\D
1}{\D\sqrt{2\pi}}\left(\begin{array}{l}f^+_i(z,r_\bot)e^{i(\Omega_i-1/2)\varphi}\\
f^-_i(z,r_\bot)e^{i(\Omega_i+1/2)\varphi}\\ig^+_i(z,r_\bot)e^{i(\Omega_i-1/2)\varphi}\\
ig^-_i(z,r_\bot)e^{i(\Omega_i+1/2)\varphi}\end{array}\right)\chi_{t_i}(t).
\end{equation}
A detailed description of the deformed RMF+BCS method can be found
in Ref. \cite{geng.03}.

In the present study, nuclei with both even and odd numbers of
protons (neutrons) need to be calculated. We adopt a simple
blocking method, in which the ground state of an odd system is
described by the wave function
\begin{equation}
\alpha^\dagger_{k_1}|\textrm{BCS}\rangle=\alpha^\dagger_{k_1}\prod_{k\ne
k_1}(u_k+v_k\alpha^\dagger_k\alpha^\dagger_{\bar{k}})|\textrm{vac}\rangle.
\end{equation}
Here, $|\textrm{vac}\rangle$ denotes the vacuum state.  The
unpaired particle sits in the level $k_1$ and blocks this level
from pairing correlations. The Pauli principle prevents this level
from participating in the scattering process of nucleons caused by
the pairing correlations. As described in Ref. \cite{ring.80}, in
the calculation of the gap, one level is ``blocked'',
\begin{equation}
\Delta_k=-\frac{\D 1}{\D 2}\sum_{k'\ne k_1>0}\frac{\D
\bar{V}_{kk'}
\Delta_{k'}}{\D\sqrt{(\varepsilon_{k'}-\lambda)^2+\Delta_{k'}^2}}.
\end{equation}
The level $k_1$ has to be excluded from the sum, because it cannot
contribute to the pairing energy. The corresponding chemical
potential, $\lambda$, is determined by
\begin{equation}
N=1+2\sum_{k\ne k_1>0}v^2_k.
\end{equation}
This blocking procedure is performed at each step of the
self-consistent iteration.

\section{Deformed ground-state proton emitters with $55\le Z\le 73$}

In the present work, we use the mass-dependent parameter set TMA
\cite{sugahara.94} for the RMF Lagrangian. Results with the
parameter set NL3 \cite{ring.97} are also presented for
comparison. For each nucleus, first the quadrupole constrained
calculation \cite{flocard.73,hirata.97} is performed to obtain all
possible ground state configurations, and then we perform the
non-constrained calculation using the quadrupole deformation
parameter of the deepest minimum of the energy curve of each
nucleus as the deformation parameter for our Harmonic-Oscillator
basis. In the case that there are several similar minima obtained
from the constrained calculation, we repeat the above-described
procedure to obtain the configuration with the lowest energy as
our final result. The calculations for the present analysis have
been performed using an expansion in 14 oscillator shells for
fermion fields and 20 shells for boson fields for isotopes with
$Z\ge 70$, i.e. Ta and Lu isotopes. While for isotopes with $Z\le
70$, 12 shells for fermion fields and 20 shells for boson fields
are used. In the latter case, we use 12 shells for fermions to
save computation time. Convergence has been tested for all the
quantities calculated here. Following Ref. \cite{gambhir.90}, we
fix $\hbar\omega_0=41A^{-1/3}$ for fermions.

Whenever the zero-range $\delta$ force is used in the BCS
framework, a cutoff procedure must be applied, i.e. the space of
the single-particle states in which the pairing interaction acts
must be truncated. This is not only to simplify the numerical
calculation but also to simulate the finite-range (more precisely,
long-range) nature of the pairing interaction in a
phenomenological way \cite{doba.96,hfb2.02}. In the present work,
the single-particle states subject to the pairing interaction are
confined to the region satisfying
\begin{equation}
\epsilon_i-\lambda\le E_\mathrm{cut},
 \end{equation}
 where $\epsilon_i$ is the single-particle energy, $\lambda$
 the Fermi energy, and $E_\mathrm{cut}=8.0$ MeV. We find that increasing
 $E_\mathrm{cut}$ from 8.0 MeV up to 16.0 MeV, followed by a
 readjustment of the pairing strength $V_0$, does not change the
 results, and therefore none of our conclusions
 will change. The same pairing strength is used for both protons and neutrons
 and is fixed by requiring that the experimental one-proton
separation energies of several selected nuclei be reproduced by
our calculations, including $^{157}$Ta, $^{147}$Tm, and
$^{111}$Cs. A slight readjustment of the pairing strength, say \%5
, does not change our conclusion here. More specifically, for
calculations with the NL3 set, we take $V_0=393.0$ MeV fm$^3$ for
$Z\ge 71$, and $V_0=420.0$ MeV fm$^3$ otherwise. For calculations
with the TMA set, we take $V_0=393.0$ MeV fm$^3$ for $Z\ge 71$,
and $V_0=375.0$ MeV fm$^3$ otherwise. We should mention that the
small difference between $V_0$ for isotopes with $Z\ge 71$ and
isotopes with $Z<71$ is partly due to the different number of
shells used in the expansion method.

In the process of proton emission, the valence proton tunnels
through the Coulomb and centrifugal barriers, and the decay
probability depends strongly on the energy of the unbound proton
and on its angular momentum. In rare-earth nuclei, the decay of
the ground state by direct proton emission competes with $\beta^+$
decay; for heavy nuclei, fission or $\alpha$-decay can also be
favored. In general, ground state proton emission is not observed
immediately beyond the proton drip line. For small values of the
proton separation energy, the width is dominated by $\beta^+$
decay. On the other hand, large separation energies result in
extremely short proton emission half-lives, which are difficult to
observe experimentally. For a typical rare-earth nucleus, the
separation energy window, in which ground state proton decay can
be directly observed, is about 0.8--1.7 MeV \cite{aberg.97}.

The Nilsson quantum numbers of the last occupied proton orbit are
taken to be the same as the dominant component in the expansion of
this wave function in terms of the axially-symmetric
Harmonic-Oscillator basis \cite{gambhir.90}. In theoretical
calculations, the spectroscopic factor of the corresponding
odd-proton orbit, $k$, is defined as the probability that this
orbit is found empty in the daughter nucleus with one less proton
\cite{lalanpa.99}. If the difference between the ground state
configurations of the mother and the daughter nuclei is ignored,
the spectroscopic factor for level $k$ is calculated as
        \begin{equation}S_k=|\langle\phi|\alpha_k|\phi_k\rangle|^2=u^2_k.\end{equation}
Here, $u^2_k$ is related to the occupation probability $v^2_k$ for
level $k$ in the daughter nucleus through the well-known relation
in the BCS formalism
         \begin{equation}u^2_k+v^2_k=1.\end{equation}
In what follows, we discuss the details of our numerical results.

\subsection{Lutetium (Z=71) and Tantalum (Z=73)}
\begin{figure}[b]
\centering
\begin{minipage}[c]{0.5\linewidth}
\includegraphics[scale=0.38]{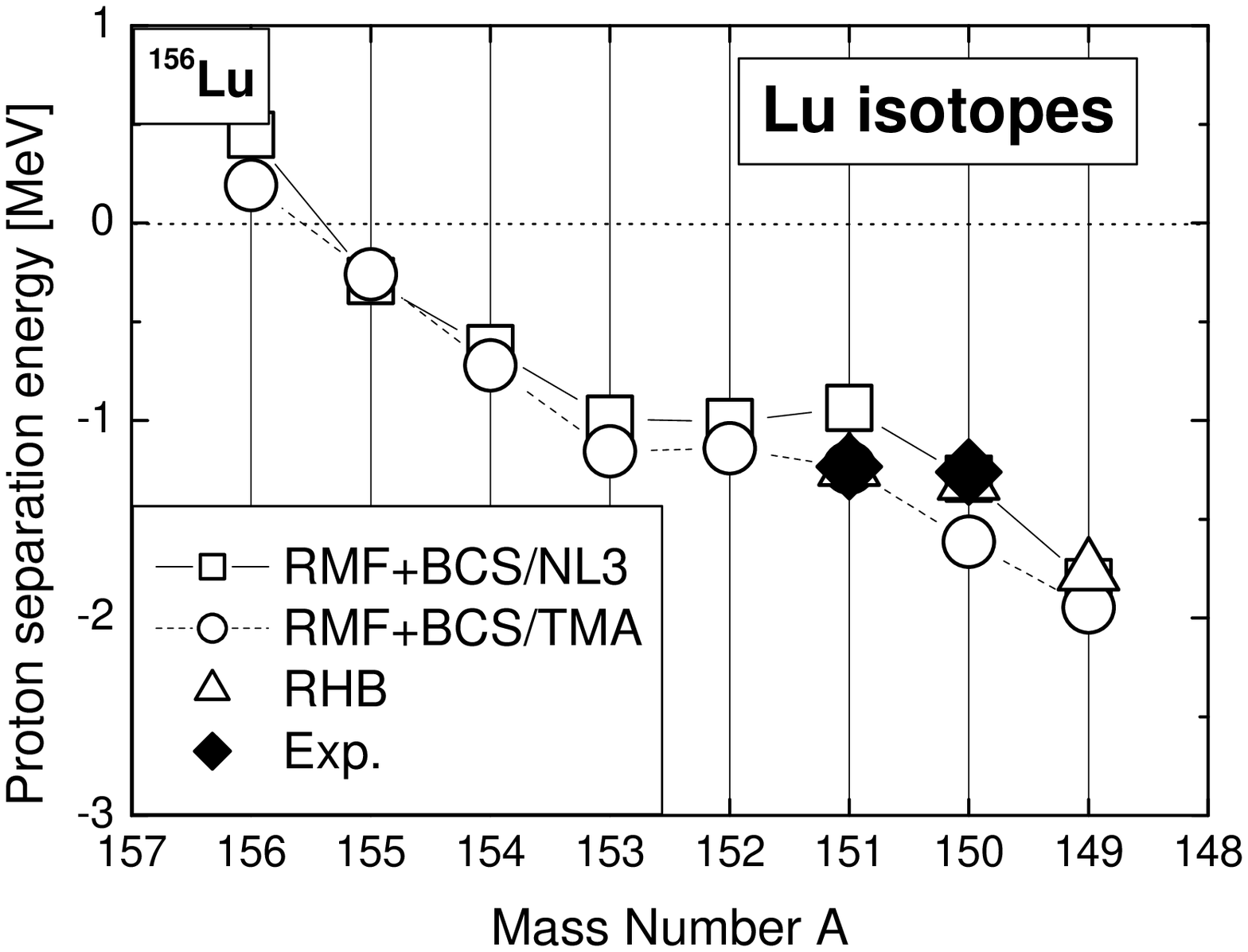}
\end{minipage}%
\begin{minipage}[c]{0.5\linewidth}
\includegraphics[scale=0.38]{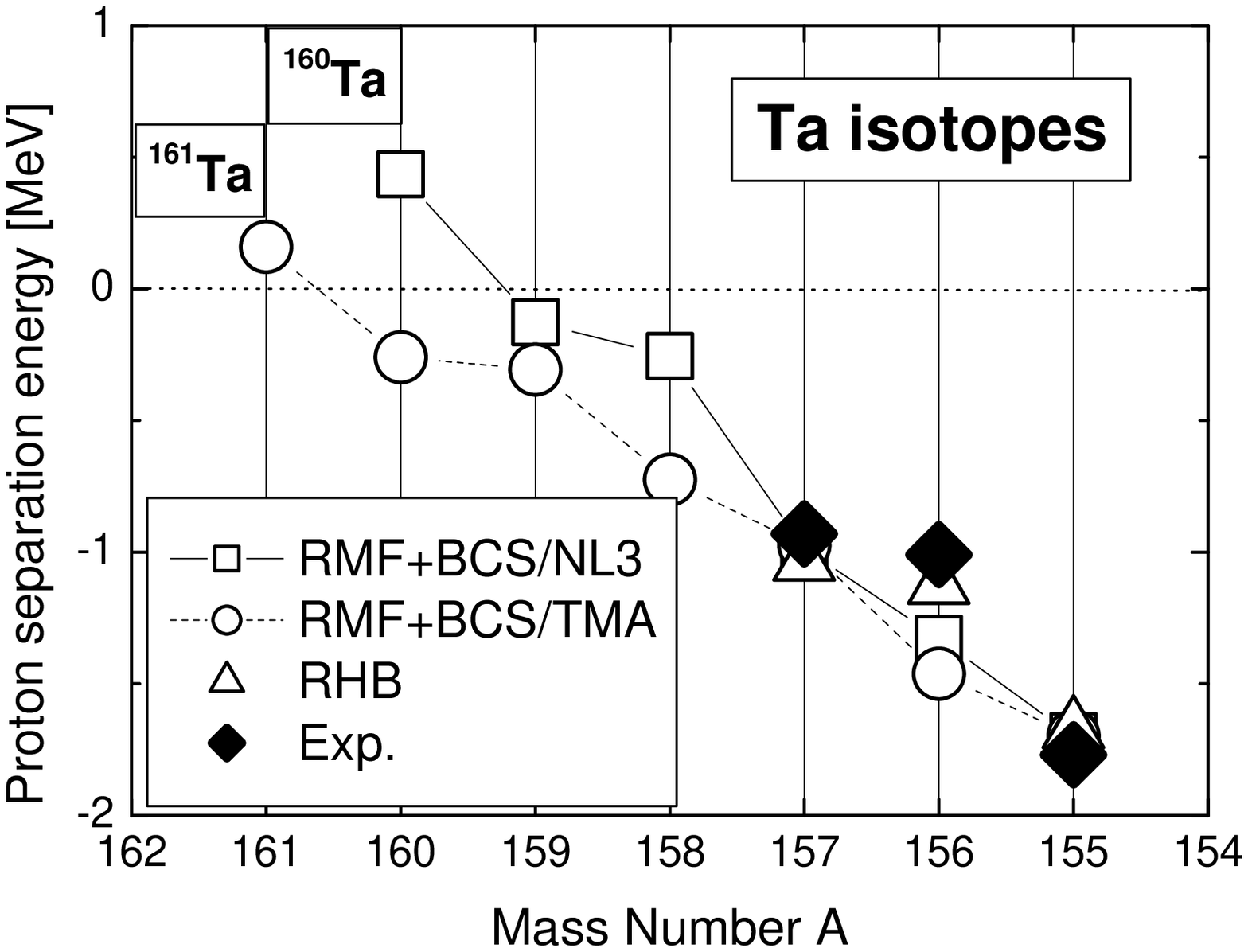}
\end{minipage}
\caption{\label{fig1.fig}The one-proton separation energies for Lu
($Z$=71) and Ta ($Z$=73) isotopes at and beyond the proton drip
line. The experimental value for the proton separation energy
corresponds to the opposite of the ground state transition energy,
$E_p$.}
\end{figure}
The one-proton separation energies for Lu and Ta isotopes are
displayed in Fig. 1 as functions of the mass number A. The
drip-line nucleus for the Ta isotopic chain is predcited to be
$^{160}$Ta by NL3 and $^{161}$Ta by TMA. While for the Lu isotopic
chain, both TMA and NL3 predict the drip-line nucleus to be
$^{156}$Lu. We also compare the calculated separation energies
with the available experimental transition energies for ground
state proton emission in $^{150}$Lu, $^{151}$Lu \cite{sell.93},
$^{155}$Ta \cite{uusi.99}, $^{156}$Ta \cite{page.92} and
$^{157}$Ta \cite{irvi.97}. The corresponding RHB results are taken
from Ref. \cite{lalaprc.99}. In Table I, we list the ground state
properties of Lu and Ta isotopes. The separation energy window,
0.8--1.7 MeV, extends to include those nuclei for which a direct
observation of ground state proton emission is in principle
possible on the basis of our calculated separation energies. Good
agreement between our calculations and both the experimental data
and the RHB results is clearly seen. However, for the odd-odd
nuclei $^{150}$Lu and $^{156}$Ta, our calculations predict values
of $S_p$ smaller than the experimental values. This may be
explained by the possible existence of a residual proton-neutron
pairing interaction. In both nuclei, an extra positive energy of
about $0.3$ MeV is needed to increase the calculated $S_p$ value
to the experimental value. We find that if we reduce the pairing
strength for odd-odd nuclei by about 5--10\% compared with
odd-even or even-even nuclei, we can fit the experimental
transition energy quite well. We see below that the same
conclusion also holds for other odd-odd nuclei. Although it has
long been known that the proton-neutron pairing is important for
the proper description of nuclei \cite{chen.67, goodman.72}, it
was believed that it only affects those proton-rich nuclei with
$N\approx Z$. However, recently \v{S}imkovic et al. \cite{sim.03}
showed that even for nuclei with $N-Z=12$, the proton-neutron
pairing cannot be ignored.

\begin{table}[t]
\setlength{\tabcolsep}{0.1em}\caption{Lu ($Z$=71) and Ta ($Z$=73)
ground state proton emitters. The results of RMF+BCS/TMA
calculations (third to sixth columns) for the one-proton
separation energy, $S_p$, the mass quadrupole deformation,
$\beta_2$, and the deformed single-particle orbit occupied by the
odd valence proton, are compared with the predictions of the RHB
model \cite{lalaprc.99,lalanpa.99} (seventh to tenth columns) and
with the experimental values, $S_p=-E_p$, where $E_p$ is the
experimental transition energy. The RMF+BCS/TMA and RHB
spectroscopic factors are displayed in the sixth and tenth
columns, respectively. All energies are in units of MeV. }
\begin{center}\label{table1}
\begin{tabular}{l@{\hspace{2ex}}c@{\hspace{2ex}}|c@{\hspace{2ex}}c@{\hspace{2ex}}c@{\hspace{2ex}}c@{\hspace{2ex}}|c@{\hspace{2ex}}c@{\hspace{2ex}}c@{\hspace{2ex}}c@{\hspace{2ex}}|c}
\hline\hline
 &&\multicolumn{4}{|c|}{RMF+BCS/TMA}&\multicolumn{4}{|c|}{RHB}&Exp.\\
 \hline
 &$N$&$S_p$&$\beta_2$&p orbital&$u^2$&$S_p$&$\beta_2$&p
 orbital&$u^2$&$S_p$ \\
 \hline
 $^{149}$Lu&78&$-1.946$&$-0.166$&$5/2^-$[512]&0.574&$-1.77$&$-0.158$&$7/2^-$[523]&0.60&\\
  $^{150}$Lu&79&$-1.613$&$-0.129$&$5/2^-$[523]&0.549&$-1.31$&$-0.153$&$7/2^-$[523]&0.61&$-1.261$(4) \cite{sell.93}\\
   $^{151}$Lu&80&$-1.239$&$-0.119$&$5/2^-$[532]&0.541&$-1.24$&$-0.151$&$7/2^-$[523]&0.58&$-1.233$(3) \cite{sell.93}\\
$^{152}$Lu&81&$-1.141$&$-0.053$&$5/2^-$[532]&0.496&&&&&\\
$^{153}$Lu&82&$-1.156$&0.002&$7/2^-$[523]&0.463&&&&&\\
$^{154}$Lu&83&$-0.720$&$-0.050$&$5/2^-$[532]&0.513&&&&&\\
\hline
$^{155}$Ta&82&$-1.698$&0.007&$9/2^-$[514]&0.374&$-1.677$&0.000&$11/2^-$&0.60&$-1.765$(10) \cite{uusi.99}\\
$^{156}$Ta&83&$-1.462$&$-0.048$&$3/2^-$[521]&0.437&$-1.129$&0.000&$3/2^+$&0.51&$-1.007$(5) \cite{page.92}\\
$^{157}$Ta&84&$-0.974$&0.036&$9/2^-$[514]&0.381&$-1.040$&0.000&$11/2^-$&0.42&$-0.927$(7) \cite{irvi.97}\\
$^{158}$Ta&85&$-0.725$&0.077&$9/2^-$[514]&0.470&&&&&\\
 \hline\hline
\end{tabular}
\end{center}
\end{table}

With regard to the deformation, both our RMF+BCS method and the
RHB model predict oblate shapes for Lu proton emitters and similar
values for the ground state quadrupole deformations. Recent
calculations by Ferreira et al. \cite{ferr.00} show that the
proton decay in $^{151}$Lu can be described very well as a decay
from a $K=5/2^-$ ground state with an oblate deformation for which
$-0.18 < \beta_2 < -0.14$. In another work \cite{ferr.01}, the
author finds that the proton decay in $^{150}$Lu is described very
well as a decay from a $K=5/2^-$ ground state with an oblate
deformation satisfying $-0.17 < \beta_2 < -0.16$. These
calculations are in reasonable agreement with our predictions (see
Table I). While the RHB method \cite{lalaprc.99} predicts that the
last occupied orbit is $7/2^-[523]$ for both nuclei. If we look
into the single-particle spectra of these two nuclei more closely,
we find that in addition to the $5/2^-$ possibilities, which are
listed in Table I, the orbits $1/2^+[400]$ and $7/2^-[523]$ are
also possibly blocked. The total energy difference between these
blocking choices are $E_{5/2}-E_{1/2}=0.061$ MeV and
$E_{5/2}-E_{7/2}=0.303$ MeV for $^{151}$Lu;
$E_{5/2}-E_{1/2}=-0.040$ MeV and $E_{5/2}-E_{7/2}=0.242$ MeV for
$^{150}$Lu, where $E$ is the binding energy for our calculations
with the TMA parameter set.

In calculations using the RHB model \cite{lalaprc.99}, Ta isotopes
are assumed to be spherical, while they are slightly deformed in
our calculations. At first sight, the RMF+BCS/TMA and RHB models
seem to predict quite different $p$--orbitals for $^{157}$Ta,
$^{156}$Ta, and $^{155}$Ta, but in fact these states are quite
close in our calculations. The theoretical predictions are
compared with the corresponding experimental assignments for the
ground state configuration and spectroscopic factor. The
experimental data for $^{155}$Ta \cite{uusi.99}, $^{156}$Ta
\cite{aberg.97} and $^{157}$Ta \cite{aberg.97} are, respectively,
the $p$--orbital $11/2^-$ with 0.58(20), the $p$--orbital $3/2^+$
with 0.67(16) and the $p$--orbital $1/2^+$ with 0.74(34). We
notice that the experimental assignments are somewhat different
both from our predictions, in which $^{155-157}$Ta are slightly
deformed, and from the RHB calculations, in which spherical shapes
are assumed. There are two possible explanations for this
disagreement. First, these states are close to each other in our
calculations and therefore it is difficult to make a clear
distinction between them. Second, deformation could slightly
change the state occupied by the odd-proton. Because the nuclei
$^{155-157}$Ta are not strongly deformed, a different combination
of deformation and the occupied state can give the same
experimentally observed values, i.e. the experimental transition
energies and decay half-lives.

\subsection{Holmium (Z=67) and Thulium (Z=69)}

\begin{figure}[t]
\centering
\begin{minipage}[c]{0.5\linewidth}
\includegraphics[scale=0.38]{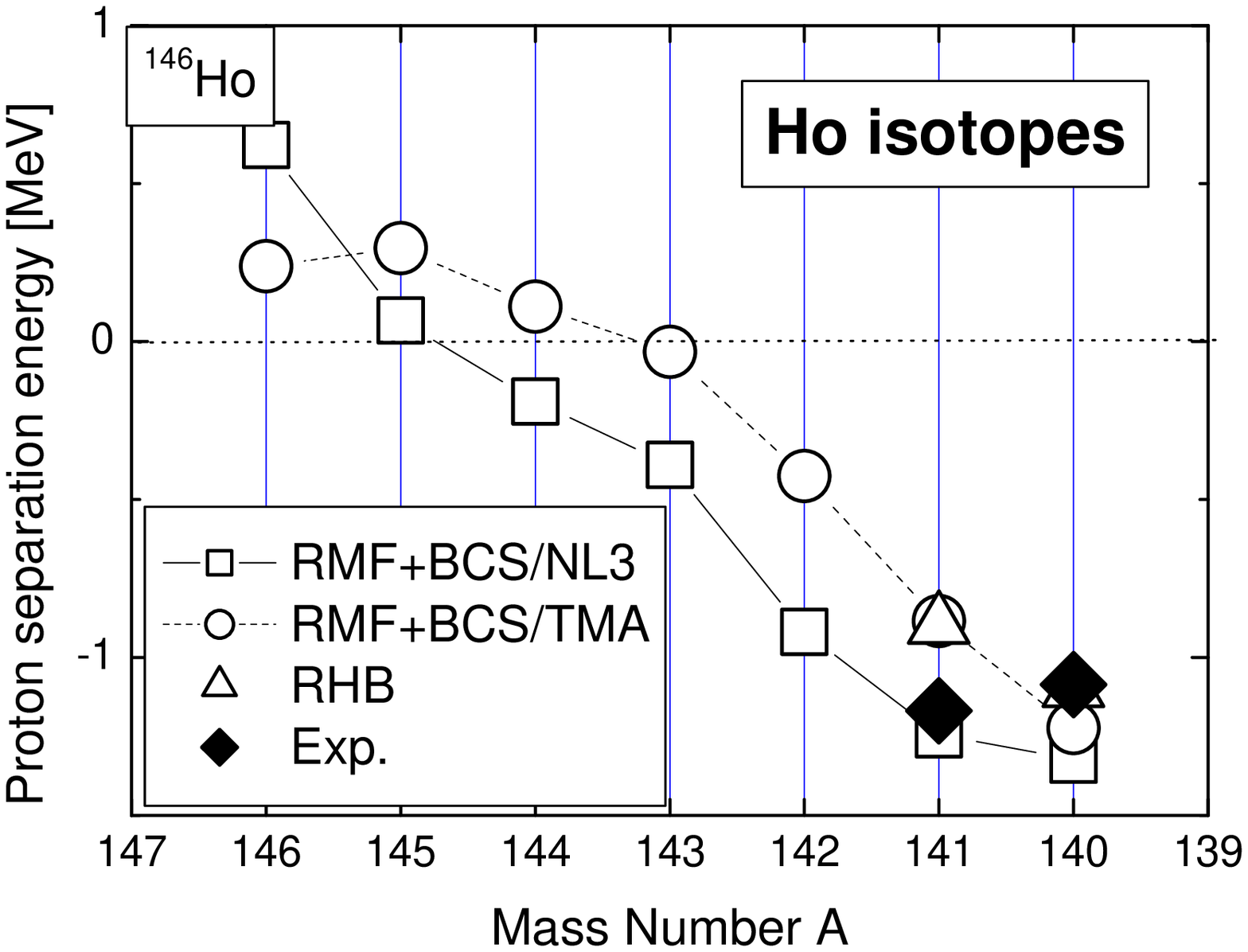}
\end{minipage}%
\begin{minipage}[c]{0.5\linewidth}
\includegraphics[scale=0.38]{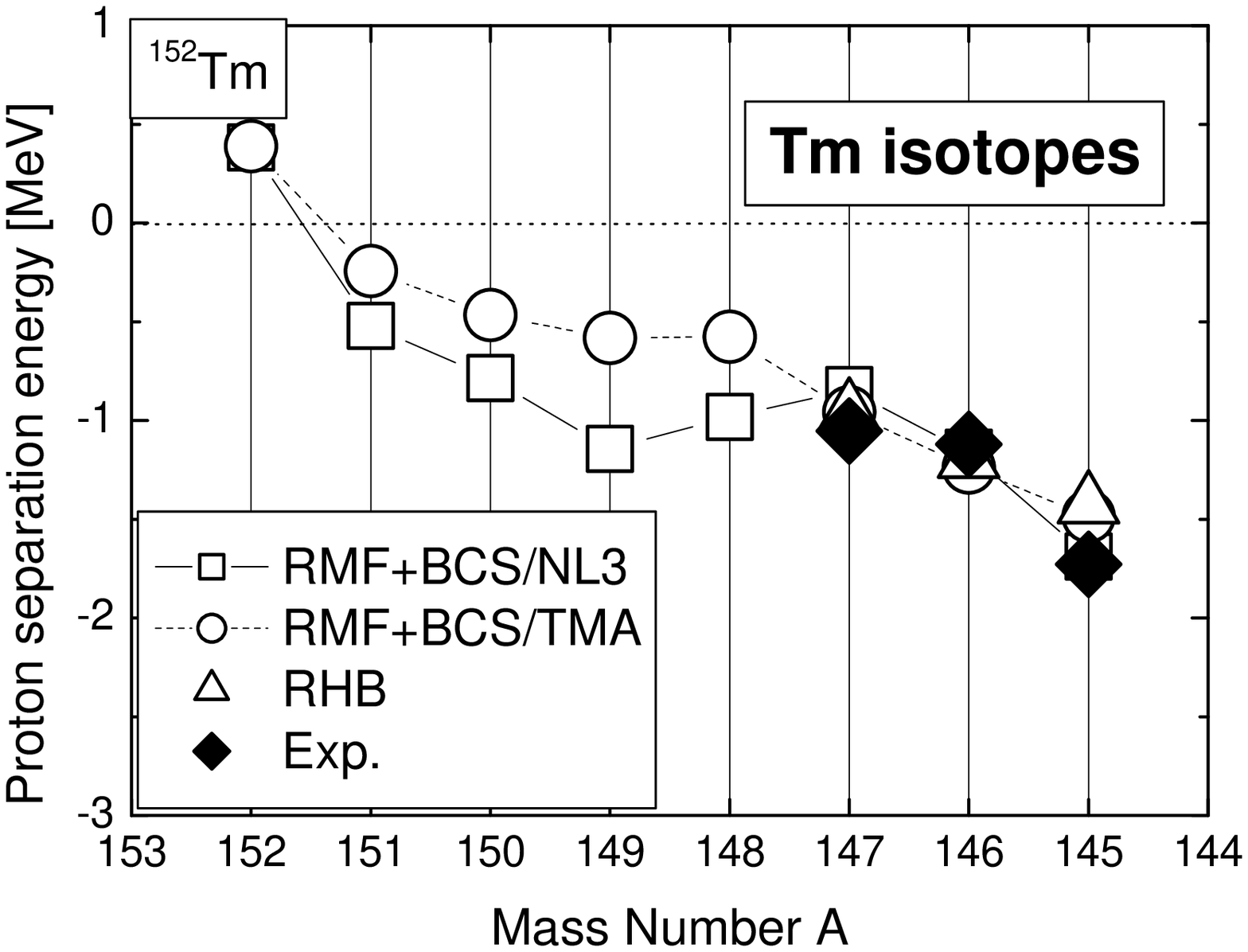}
\end{minipage}
\caption{\label{fig2.fig}The same as Fig. \ref{fig1.fig}, but for
Ho ($Z$=67) and Tm ($Z$=69) isotopes.}
\end{figure}
\begin{table}[b]
\setlength{\tabcolsep}{0.1em}\caption{The same as Table
\ref{table1}, but for Ho ($Z$=67) and Tm ($Z$=69) isotopes.}
\begin{center}\label{table2}
\begin{tabular}{l@{\hspace{2ex}}c@{\hspace{2ex}}|c@{\hspace{2ex}}c@{\hspace{2ex}}c@{\hspace{2ex}}c@{\hspace{2ex}}|c@{\hspace{2ex}}c@{\hspace{2ex}}c@{\hspace{2ex}}c@{\hspace{2ex}}|c}
\hline\hline
 &&\multicolumn{4}{|c|}{RMF+BCS/TMA}&\multicolumn{4}{|c|}{RHB}&Exp.\\
 \hline
 &$N$&$S_p$&$\beta_2$&p orbital&$u^2$&$S_p$&$\beta_2$&p
 orbital&$u^2$&$S_p$\\
 \hline
 $^{140}$Ho&73&$-1.223$&0.341&$7/2^-[523]$&0.592&$-1.10$&0.31&$7/2^-$[523]&0.61&$-1.086$(10) \cite{rykac.99}\\
 $^{141}$Ho&74&$-0.885$&0.339&$7/2^-[523]$&0.594&$-0.90$&0.32&$7/2^-$[523]&0.64&$-1.169$(8) \cite{david.98}\\
  \hline
   $^{145}$Tm&76&$-1.458$&$-0.220$&$7/2^-[523]$&0.458&$-1.43$&0.23&$7/2^-$[523]&0.47&$-1.728$(10) \cite{batc.98}\\
 $^{146}$Tm&77&$-1.238$&$-0.207$&$7/2^-[523]$&0.460&$-1.20$&$-0.21$&$7/2^-$[523]&0.50&$-1.120$(10) \cite{livi.93}\\
  $^{147}$Tm&78&$-0.958$&$-0.186$&$7/2^-[523]$&0.467&$-0.96$&$-0.19$&$7/2^-$[523]&0.55&$-1.054$(19) \cite{sell.93}\\
 \hline\hline
\end{tabular}
\end{center}
\end{table}
In Fig. 2, we plot the one-proton separation energies for Ho and
Tm isotopes as functions of the mass number A. The available
experimental data for $^{147}$Tm \cite{sell.93}, $^{146}$Tm
\cite{livi.93}, $^{145}$Tm \cite{batc.98}, $^{140}$Ho
\cite{rykac.99} and $^{141}$Ho \cite{david.98}, together with the
results of the RHB calculations \cite{lalanpa.99}, are also shown
for comparison. The predicted drip-line nucleus for the Tm
isotopic chain is $^{152}$Tm, while for the Ho isotopic chain, NL3
predicts $^{146}$Ho and TMA predicts $^{144}$Ho. From Fig.
\ref{fig2.fig}, we can see that both RMF+BCS/TMA and RHB predict a
smaller transition energy for $^{141}$Ho, while RMF+BCS/NL3
predicts a value of $S_p$ that is closer to the experimental
value. Nevertheless, all theoretical calculations fail to predict
the increase of the one-proton separation energy from $^{141}$Ho
to $^{140}$Ho. This can be corrected by reducing the pairing
strength by a few percent, as discussed above. However, this is
not our aim here. It is quite clear that, even after changing the
pairing strength, we cannot account for the trend from $^{141}$Ho
to $^{140}$Ho, which has been attributed to the lack of a residual
proton-neutron interaction in our above analysis. In Table II, we
list the properties of the Ho and Tm proton emitters. We note that
our RMF+BCS/TMA calculations predict an oblate shape for
$^{145}$Tm, while the RHB method predicts a prolate shape. In
fact, in our constrained calculation, both prolate and oblate
shapes are possible for $^{145}$Tm and $^{144}$Er. This is
illustrated in Fig. 3, where we plot the binding energy per
particle for $^{145}$Tm and $^{144}$Er as functions of the
quadrupole deformation parameter, $\beta_{2m}$. The binding
energies are obtained from the RMF+BCS/TMA calculations performed
by imposing a quadratic constraint on the quadrupole moment. We
find two similar minima around $\beta_2\approx -0.22$ and
$\beta_2\approx 0.25$ for both $^{145}$Tm and $^{144}$Er. More
specifically, for $^{145}$Tm the two minima are $E/A=-7.906$ at
$\beta_{2m}=0.287$ and $E/A=-7.903$ at $\beta_{2m}=-0.211$, while
for $^{144}$Er they are $E/A=-7.767$ at $\beta_{2m}=0.303$ and
$E/A=-7.768$ at $\beta_{2m}=-0.196$. The reason that we choose the
oblate shape instead of the prolate shape for $^{145}$Tm is
two-fold. First, we would like to choose the same shape for both
$^{145}$Tm and $^{144}$Er. Second, the oblate shape for $^{145}$Tm
is more consistent with its neighbors (also see Fig. 7). We note
that, unlike Lu and Ta isotopes, here both RMF+BCS/TMA and RHB
predict the same last occupied odd-proton state for Ho and Tm
isotopes (also see Table II).

\begin{figure}[t]
\centering
\includegraphics[scale=0.6]{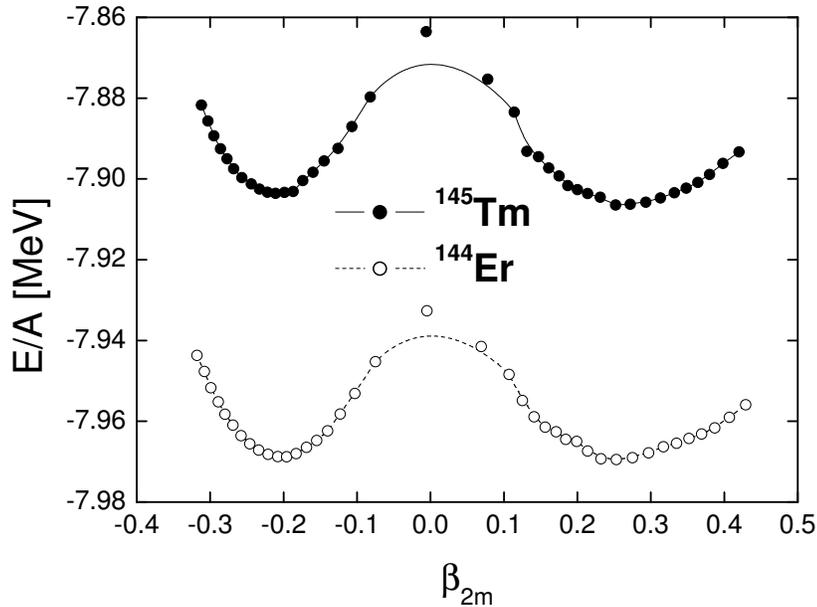}
\caption{\label{fig3.fig}Binding energy per particle, $E/A$, for
$^{145}$Tm and $^{144}$Er, plotted as functions of the mass
quadrupole deformation parameter, $\beta_{2m}$.}
\end{figure}
\subsection{Europium (Z=63) and Terbium (Z=65)}

\begin{figure}[t]
\centering
\begin{minipage}[c]{0.5\linewidth}
\includegraphics[scale=0.38]{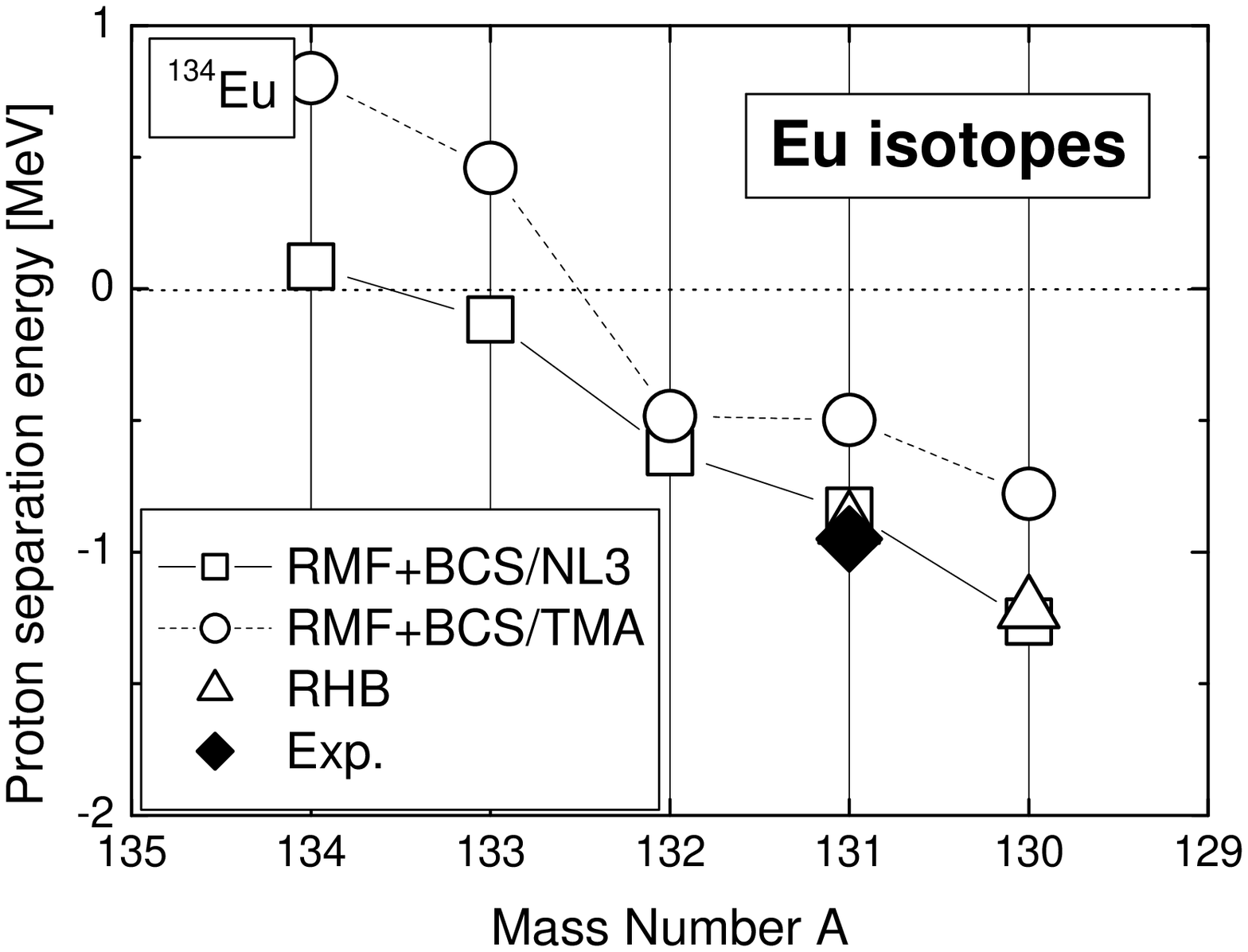}
\end{minipage}%
\begin{minipage}[c]{0.5\linewidth}
\includegraphics[scale=0.38]{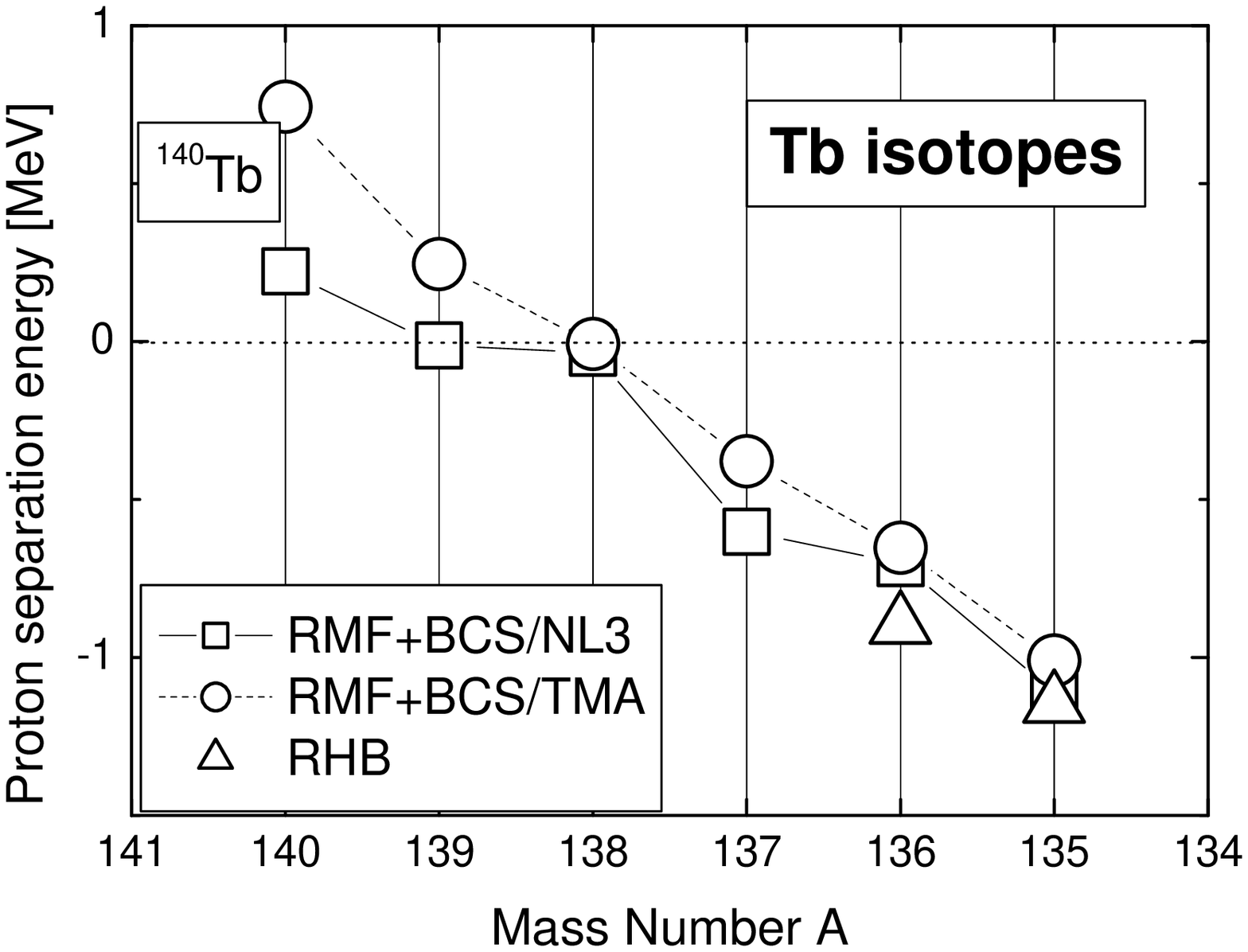}
\end{minipage}
\caption{\label{fig4.fig}The same as Fig. \ref{fig1.fig}, but for
Eu ($Z$=63) and Tb ($Z$=65) isotopes.}
\end{figure}
In Fig. 4, we plot the one-proton separation energies for Tb and
Eu isotopes. The results of the RHB model \cite{lalanpa.99} and
the only available experimental data for $^{131}$Eu
\cite{david.98} are also shown. The drip-line nucleus for the Tb
isotopic chain is predicted to be $^{140}$Tb by NL3 and to be
$^{139}$Tb by TMA. While for the Eu isotopic chain, the drip-line
nucleus is predicted to be $^{133}$Eu by TMA and $^{134}$Eu by
NL3. In this region, only one proton emitter, $^{131}$Eu, has been
reported. However, based on our calculations, there are three
other possible proton emitters, $^{130}$Eu, $^{135}$Tb and
$^{136}$Tb. We list their properties in Table \ref{table3}. We
notice that the experimental transition energy for $^{131}$Eu is
0.950(8) MeV, while our RMF+BCS/TMA calculations predict 0.498
MeV. This can be corrected by adjusting the pairing strength
slightly. Considering this correction, we see that RHB and RMF+BCS
predict similar proton emitters for Tb and Eu isotopes. As we have
seen from the results presented to this point, the description of
proton emitters is quite sensitive to the pairing strength. We
hope that further experimental measurements of proton emitters in
the entire region can help us understand more about the pairing
interaction in exotic nuclei.

\begin{table}[h]
\setlength{\tabcolsep}{0.1em}\caption{The same as Table
\ref{table1}, but for Eu ($Z$=63) and Tb ($Z$=65) isotopes.}
\begin{center}\label{table3}
\begin{tabular}{l@{\hspace{2ex}}c@{\hspace{2ex}}|c@{\hspace{2ex}}c@{\hspace{2ex}}c@{\hspace{2ex}}c@{\hspace{2ex}}|c@{\hspace{2ex}}c@{\hspace{2ex}}c@{\hspace{2ex}}c@{\hspace{2ex}}|c}
\hline\hline
 &&\multicolumn{4}{|c|}{RMF+BCS/TMA}&\multicolumn{4}{|c|}{RHB}&Exp.\\
 \hline
 &$N$&$S_p$&$\beta_2$&p orbital&$u^2$&$S_p$&$\beta_2$&p
 orbital&$u^2$&$S_p$\\
 \hline
 $^{130}$Eu&67&$-0.778$&0.420&$5/2^+[413]$&0.464&$-1.22$&0.34&$5/2^-$[532]&0.44&\\
 $^{131}$Eu&78&$-0.498$&0.411&$5/2^+[413]$&0.323&$-0.90$&0.35&$5/2^+$[413]&0.44&$-0.950$(8) \cite{david.98}\\
  \hline
   $^{135}$Tb&70&$-1.010$&0.365&$3/2^+[411]$&0.879&$-1.15$&0.34&$3/2^+$[411]&0.62&\\
 $^{136}$Tb&71&$-0.653$&0.374&$3/2^+[411]$&0.862&$-0.90$&0.32&$3/2^+$[411]&0.65&\\
 \hline\hline
\end{tabular}
\end{center}
\end{table}

\subsection{Praseodymium (Z=59) and Promethium (Z=61)}
In Fig. 5, the one-proton separation energies for Pm and Pr
isotopes are plotted as functions of the mass number A, together
with the results of the RHB model \cite{lalanpa.99}. The drip-line
nuclei are predicted to be $^{128}$Pm and $^{125}$Pr by both TMA
and NL3. No experimental proton emitters have been reported for
these two isotopic chains. Based on our calculations, possible
proton emitters in this region are $^{119,120}$Pr and
$^{124,125}$Pm. The properties of these nuclei are listed in Table
IV.

\begin{figure}[t]
\centering
\begin{minipage}[c]{0.5\linewidth}
\includegraphics[scale=0.38]{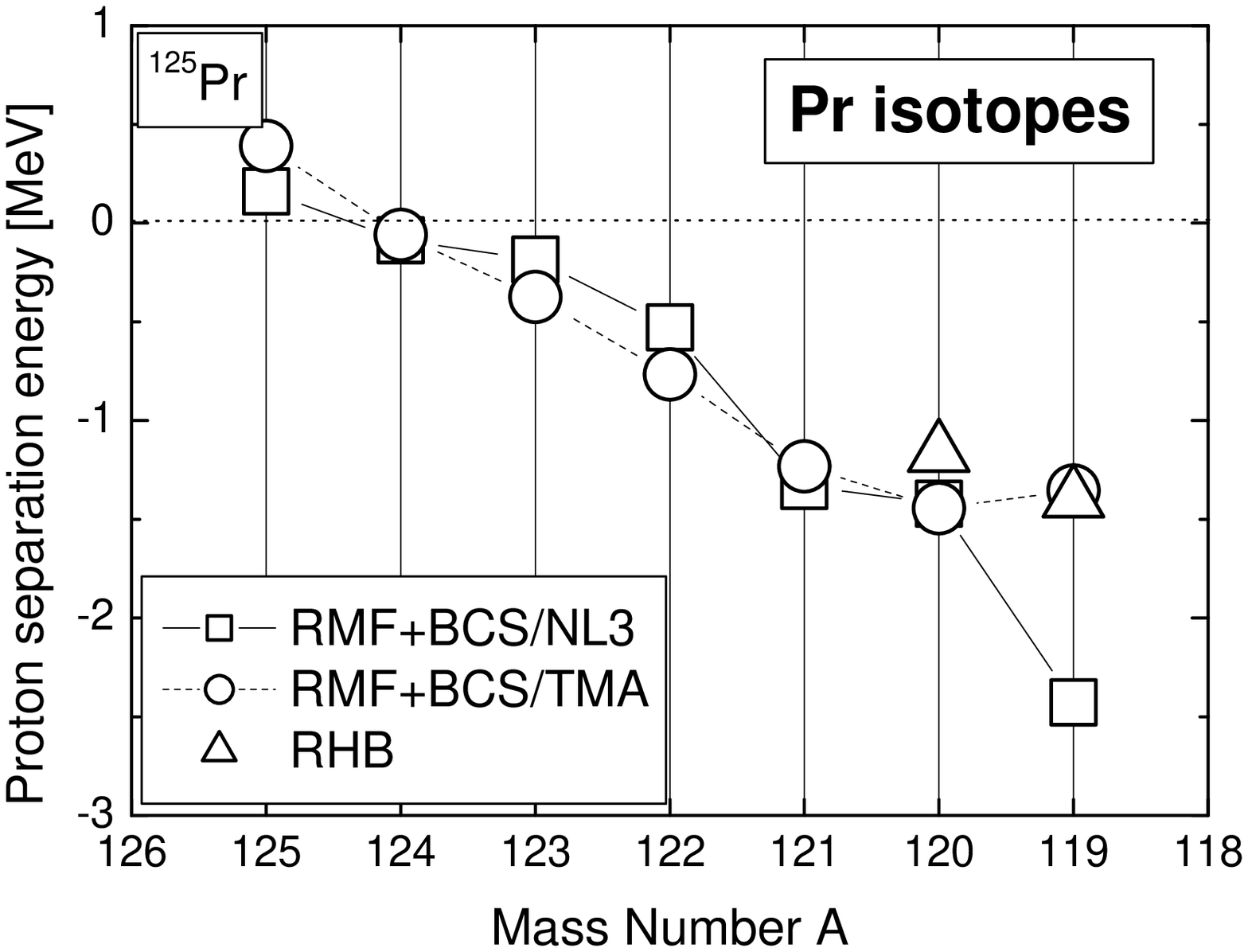}
\end{minipage}%
\begin{minipage}[c]{0.5\linewidth}
\includegraphics[scale=0.38]{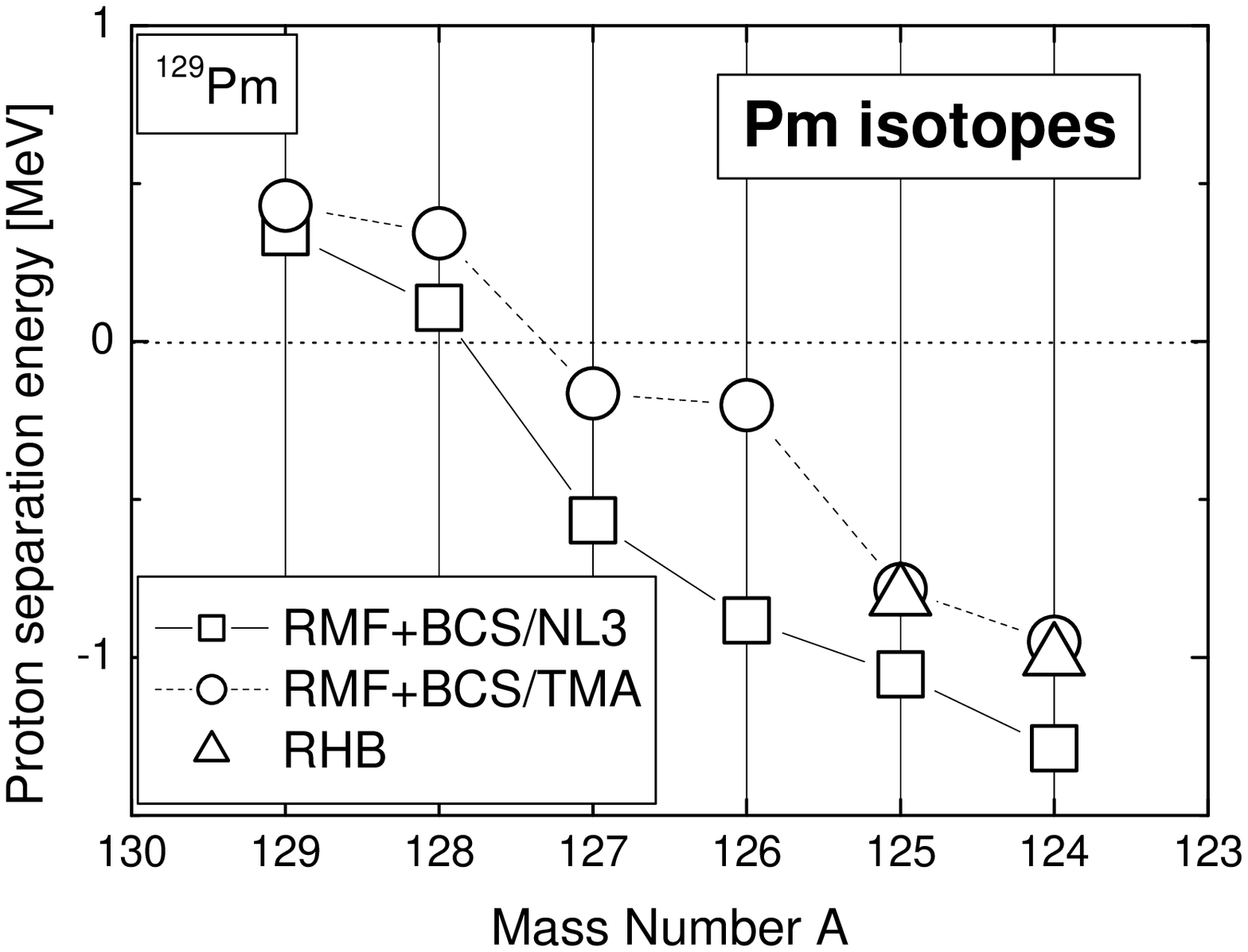}
\end{minipage}
\caption{\label{fig5.fig}The same as Fig. \ref{fig1.fig}, but for
Pr ($Z$=59) and Pm ($Z$=61) isotopes.}
\end{figure}

\begin{figure}[t]
\centering
\begin{minipage}[c]{0.5\linewidth}
\includegraphics[scale=0.38]{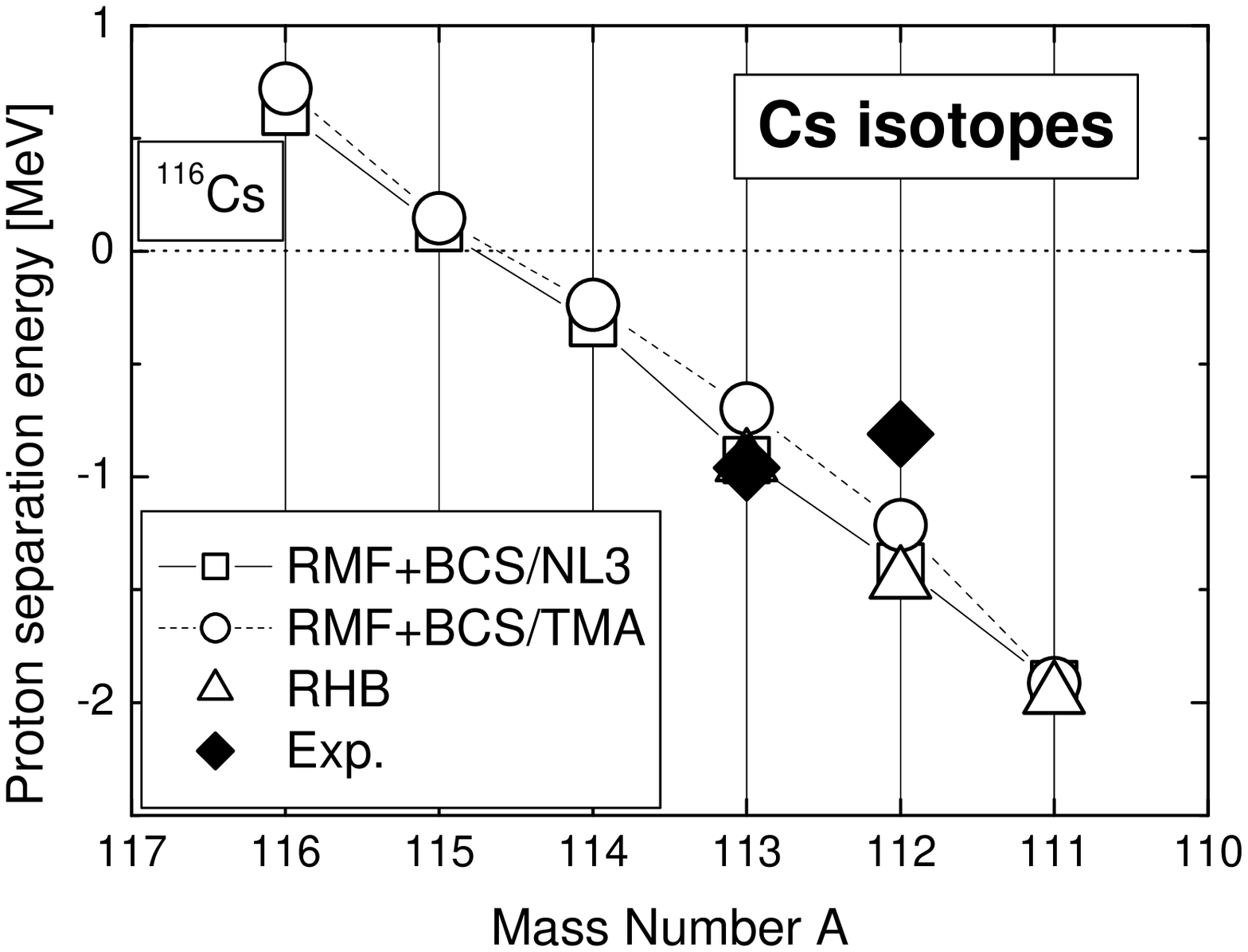}
\end{minipage}%
\begin{minipage}[c]{0.5\linewidth}
\includegraphics[scale=0.38]{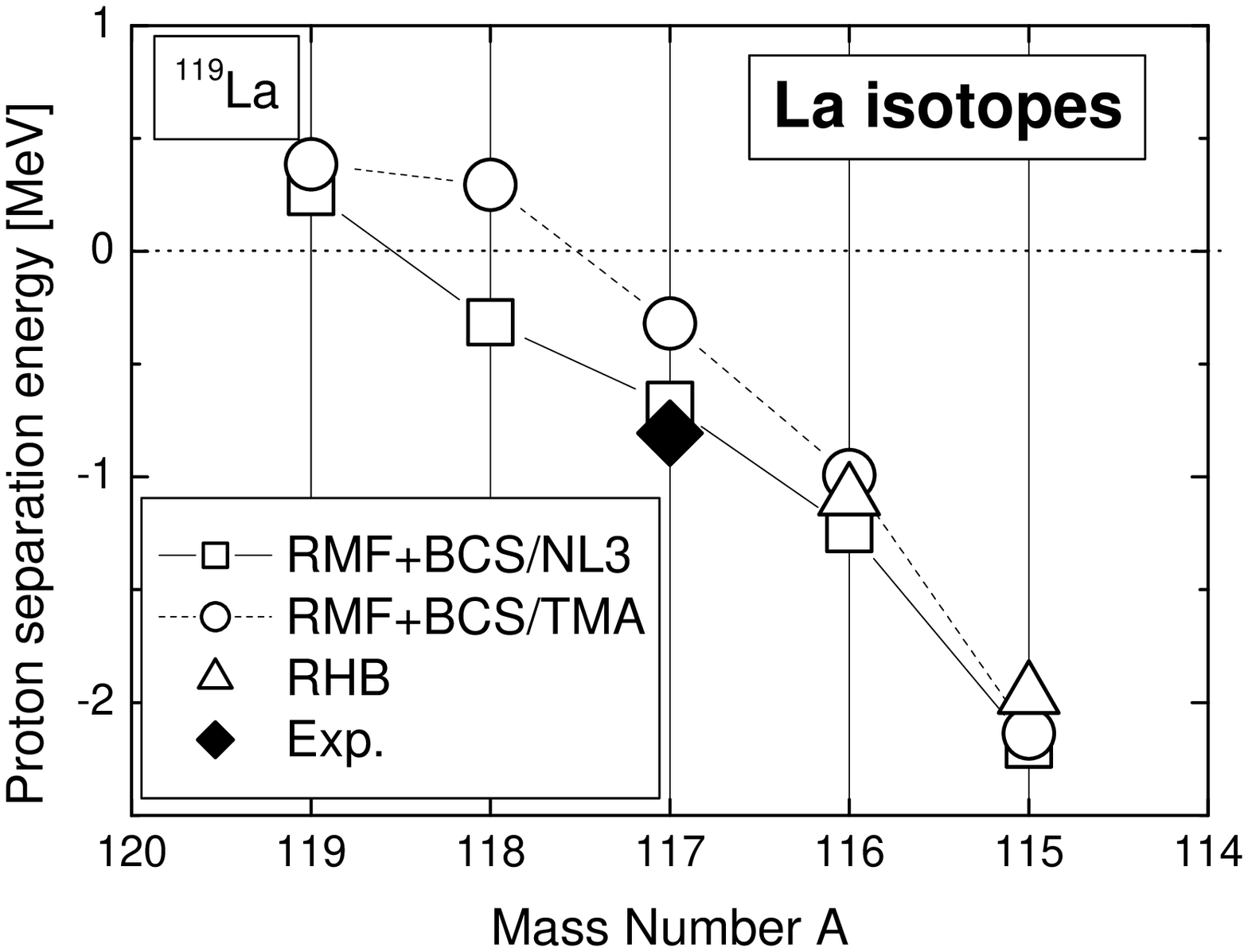}
\end{minipage}
\caption{\label{fig6.fig}The same as Fig. \ref{fig1.fig}, but for
Cs ($Z$=55) and La ($Z$=57) isotopes.}
\end{figure}
\subsection{Cesium (Z=55) and Lanthanum (Z=57)}

In Fig. \ref{fig6.fig}, we plot the one-proton separation energies
for Cs and La isotopes. The results of the RHB model
\cite{lalanpa.99} are also shown for comparison. The drip-line
nuclei are predicted to be $^{115}$Cs and $^{118}$La by TMA,
$^{115}$Cs and $^{119}$La by NL3. In Fig. 6, we see that the
latest reported proton emitter, $^{117}$La \cite{mahm.01,
sora.01}, is reproduced quite well by RMF+BCS/NL3 calculations.
Both RHB and RMF+BCS fail to reproduce the experimental data for
$^{112}$Cs. As mentioned above, $^{112}$Cs has an odd number of
protons and an odd number of neutrons. Since $N-Z$ is only 2, we
expect a relatively strong interaction between the odd proton and
the odd neutron. Compared with an odd-even system, the additional
interaction increases the binding energy somewhat. Because the two
mean-field models, RHB and RMF+BCS, do not include any residual
proton-neutron interaction, they cannot reproduce the inversion of
separation energies. Also, it is suggested in Ref.
\cite{lalanpa.99} that such an additional interaction could be
represented by a surface delta-function interaction. According to
our calculations, the remaining possible proton emitters in the La
isotopic chain are $^{115}$La and $^{116}$La, as listed in Table
V.

\begin{table}[t]
\setlength{\tabcolsep}{0.4em}\caption{The same as Table
\ref{table1}, but for Pr ($Z$=59) and Pm ($Z$=61) isotopes.}
\begin{center}\label{table4}
\begin{tabular}{l@{\hspace{2ex}}c@{\hspace{2ex}}|c@{\hspace{2ex}}c@{\hspace{2ex}}c@{\hspace{2ex}}c@{\hspace{2ex}}|c@{\hspace{2ex}}c@{\hspace{2ex}}c@{\hspace{2ex}}c@{\hspace{2ex}}|c}
\hline\hline
 &&\multicolumn{4}{|c|}{RMF+BCS/TMA}&\multicolumn{4}{|c|}{RHB}&Exp.\\
 \hline
 &$N$&$S_p$&$\beta_2$&p orbital&$u^2$&$S_p$&$\beta_2$&p
 orbital&$u^2$&$S_p$\\
 \hline
 $^{119}$Pr&60&$-1.354$&0.381&$3/2^-$[541]&0.436&$-1.40$&0.32&$3/2^-$[541]&0.39&\\
 $^{120}$Pr&61&$-1.445$&0.442&$9/2^+$[404]&0.107&$-1.17$&0.33&$3/2^-$[541]&0.33&\\
  \hline
   $^{124}$Pm&63&$-0.950$&0.422&$5/2^-$[532]&0.789&$-1.00$&0.35&$5/2^-$[532]&0.72&\\
 $^{125}$Pm&64&$-0.783$&0.412&$5/2^-$[532]&0.800&$-0.81$&0.35&$5/2^-$[532]&0.74&\\
 \hline\hline
\end{tabular}
\end{center}
\end{table}

\begin{table}[t]
\setlength{\tabcolsep}{0.2em}\caption{The same as Table
\ref{table1}, but for Cs ($Z$=55) and La ($Z$=57) isotopes.}
\begin{center}\label{table5}
\begin{tabular}{l@{\hspace{2ex}}c@{\hspace{2ex}}|c@{\hspace{2ex}}c@{\hspace{2ex}}c@{\hspace{2ex}}c@{\hspace{2ex}}|c@{\hspace{2ex}}c@{\hspace{2ex}}c@{\hspace{2ex}}c@{\hspace{2ex}}|c}
\hline\hline
 &&\multicolumn{4}{|c|}{RMF+BCS/TMA}&\multicolumn{4}{|c|}{RHB}&Exp.\\
 \hline
 &$N$&$S_p$&$\beta_2$&p orbital&$u^2$&$S_p$&$\beta_2$&p
 orbital&$u^2$&$S_p$\\
 \hline
 $^{111}$Cs&56&$-1.913$&0.206&$1/2^+$[420]&0.949&$-1.97$&0.20&$1/2^+$[420]&0.74&\\
 $^{112}$Cs&57&$-1.213$&0.171&$5/2^+$[413]&0.906&$-1.46$&0.20&$1/2^+$[420]&0.74&$-0.807$(7) \cite{page.94}\\
  $^{113}$Cs&58&$-0.697$&0.222&$1/2^+$[420]&0.94&$-0.94$&0.21&$1/2^+$[420]&0.73&$-0.9593$(37) \cite{batc.98}\\
  \hline
   $^{115}$La&58&$-2.136$&0.281&$1/2^-$[550]&0.994&$-1.97$&0.26&$1/2^+$[420]&0.20&\\
    $^{116}$La&59&$-0.992$&0.357&$3/2^-$[541]&0.8&$-1.09$&0.30&$3/2^-$[541]&0.73&\\
     $^{117}$La&59&$-0.320$&0.343&$1/2^+$[420]&0.58&&&&&$-0.806$(5) \cite{mahm.01}\\

 \hline\hline
\end{tabular}
\end{center}
\end{table}

\subsection{Quadrupole deformation}

The calculated mass quadrupole deformation parameter,
$\beta_{2m}$, for odd-$Z$ and even-$Z$ nuclei with $54\le Z \le
73$ at and beyond the proton drip line are plotted in Fig. 7 as
functions of the neutron number N. While prolate deformations
($0.15 \le \beta_{2m} \le 0.25$) are calculated for Cs isotopes,
the proton-rich isotopes of La, Pr, Pm, Eu and Tb are strongly
deformed ($0.35\le \beta_2\le 0.45$). By increasing the number of
neutrons, Ho isotopes are caused to display a transition from
prolate to oblate shapes, while most Tm nuclei have oblate shapes.
Lu and Ta isotopes exhibit a transition from oblate to prolate
shapes. The absolute value of $\beta_{2m}$ decreases as the
neutron number approaches the conventional magic number, $N=82$.

\begin{figure}[t]
\centering
\includegraphics[scale=0.7]{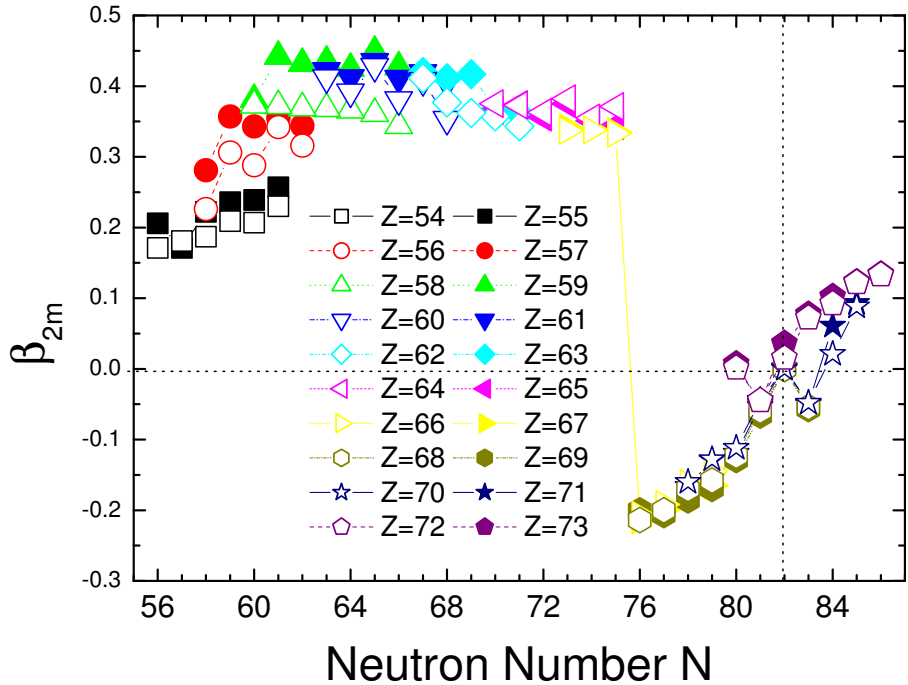}
\caption{\label{fig7.fig}Ground state quadrupole deformations for
proton drip-line nuclei with $54\le Z \le 73$ as functions of the
neutron number, $N$, at and beyond the proton drip line.}
\end{figure}

For stable even-even nuclei, the information concerning the
deformation parameter can be derived from measurements of the
$B(E2)\uparrow$ values. The $B(E2)\uparrow$ values are basic
experimental quantities that do not depend on the nuclear model.
Assuming a uniform charge distribution out to a distance
$R(\theta,\phi)$ and zero charge beyond, $\beta_{2p}$ is related
to $B(E2)\uparrow$ by
 \begin{equation}
 \beta_{2p}=(4\pi/3Z R^2_0)[B(E2)\uparrow/e^2]^{1/2},
 \end{equation}
where $R_0$ has been taken to be $1.2A^{1/3}$ fm and
$B(E2)\uparrow$ is in units of $e^2b^2$. Unfortunately, we do not
have much knowledge about the $B(E2)\uparrow$ values for proton
drip-line nuclei. The two available experimental results
\cite{raman.01} in the region are $\beta_{2p}=0.221(7)$ for
$^{114}$Xe and $\beta_{2p}=0.385(48)$ for $^{124}$Ce, which are in
reasonable agreement with our calculations: $\beta_{2p}=0.199$ for
$^{114}$Xe and $\beta_{2p}=0.344$ for $^{124}$Ce (also see Fig.
7).

Another source of knowledge for the deformation of proton
drip-line nuclei can be obtained from analysis of the properties
of proton emitters, as demonstrated in Refs.
\cite{magl.98,magl.99, ferr.00,magl.00,ferr.01}. In those works,
to simplify the calculations the assumption that the parent and
daughter nuclei have the same deformation is used. Our
calculations show that for Ho, Tm, Lu and Ta isotopes, the parent
and daughter nuclei have almost the same deformation. However, for
the other six isotopes, an absolute deviation of 0.02--0.09 is
found to exist. To what extent such a deviation can influence the
conclusions drawn from those calculations \cite{magl.98,magl.99,
ferr.00,magl.00,ferr.01} needs to be studied in more detail.

\section{Conclusion}

In the present work, the deformed RMF+BCS model has been used to
study the ground state properties of the proton drip-line nuclei
with $55\le Z\le 73$, including the location of the proton drip
line, the ground state quadrupole deformation, the one-proton
separation energy, the deformed single-particle orbital occupied
by the odd valence proton, and the corresponding spectroscopic
factor. The RMF+BCS model reproduces the available experimental
data reasonably well, except in the case of odd-odd nuclei. The
results also agree well with those of the RHB method. The
systematic discrepancies for the one-proton separation energies of
the odd-odd proton-rich nuclei may be attributed to the lack of a
residual proton-neutron pairing in both mean field models. It will
be very interesting to include a residual proton-neutron pairing
into the relativistic mean field model through a generalized BCS
method in a future work.

In conclusion, we have found that the RMF+BCS model can describe
the ground state properties of proton emitters as well as the more
complicated RHB model. It is also found that the state-dependent
BCS method with a zero-range $\delta$-force is valid not only in
the stable region but also in the proton drip line. To summarize,
the drip-line nuclei predicted by the RMF+BCS/TMA calculations are
$^{161}$Ta, $^{156}$Lu, $^{152}$Tm, $^{144}$Ho, $^{139}$Tb,
$^{133}$Eu, $^{129}$Pm, $^{125}$Pr, $^{118}$La and $^{115}$Cs.
Based on our calculations, possible proton emitters that could be
detected in future experiments are $^{149}$Lu, $^{152}$Lu,
$^{153}$Lu and $^{154}$Lu for the Lu isotopic chain, $^{158}$Ta
for the Ta isotopic chain, $^{130}$Eu for the Eu isotopic chain,
$^{135}$Tb and $^{136}$Tb for the Tb isotopic chain, $^{119}$Pr
and $^{120}$Pr for the Pr isotopic chain, $^{124}$Pm and
$^{125}$Pm for the Pm isotopic chain, $^{111}$Cs for the Cs
isotopic chain, and $^{115}$La and $^{116}$La for the La isotopic
chain.

\section{Acknowledgments}

L. S. Geng is grateful for the Monkasho Fellowship that supported
his stay at Research Center for Nuclear Physics, where this work
was performed .

\end{document}